

\input jnl
\input reforder
\input defs
\input eqnorder

\rightline{NSF-ITP-95-43}

\title Determinism plus chance in random matrix theory
\vskip .2in
\author A. Zee

\affil
Institute for Theoretical Physics
University of California
Santa Barbara, CA 93106-4030 USA

\abstract
{We study Hamiltonians consisting of a deterministic term plus a random
term. Using a diagrammatic approach and introducing the concept of ``gluon
connectedness," we calculate the density of energy levels for a wide class
of probability distributions governing the random term, thus generalizing a
result obtained recently by Br\'ezin, Hikami, and Zee. The method used here
may be applied to a broad class of problems involving random matrices.}

\vskip .3in

Some four decades ago, Wigner\refto{WIG} proposed
studying the
distribution of energy levels of a random Hamiltonian given
by
$$
H=\varphi
\eqno(random)
$$
where $\varphi$ is an $N$ by $N$ hermitean matrix taken
from the distribution
$$
P(\varphi)={1 \over Z}{e^{{-N}
trV(\varphi)}}.\eqno(distribution)
$$ with $Z$ fixed by $\int d
\varphi
P(\varphi)=1$.
This problem has been studied intensively by Dyson, Mehta, and others over the
years.\refto{POR,MEH,JER}
Two years ago, Br\'ezin and Zee discovered that, remarkably, while the
density of eigenvalues depends\refto{BIPZ} on $V$,
the correlation between the density of eigenvalues,
when suitably scaled, is
independent\refto{BZ1} of $V$. This universality has been clarified and
extended by other authors,\refto{bee, eyn, for} studied
numerically,\refto{koba} and furthermore, shown to hold even when the
distribution \(distribution) is generalized to a
much broader class of distributions.\refto{BZ2} We expect that the
discussion to be given below will hold also for this broader class of
distributions, but for the sake of simplicity we will not work this
through here.

In recent work,\refto{BZ3, BHZ}  Br\'ezin and Zee have generalized
this Wigner problem to the case of a Hamiltonian given by
the sum of a deterministic term and a random term
$$
H= H_0 + \varphi
\eqno(det)
$$
Here $H_0$ is a diagonal matrix
with diagonal
elements $\epsilon_i$, $i=1,2,...N$, and $\varphi$ a random
matrix taken from the ensemble \(distribution). For the
Gaussian case, namely with
$V(\varphi)={1\over2}\varphi^2$, Pastur\refto{PAS} has
long ago determined the density of eigenvalues. The work described in
\Ref{BZ3} went beyond Pastur's work in that the
correlation function between the density of eigenvalues in the Gaussian
case was also determined. More recently, in a work with Br\'ezin and
Hikami,\refto{BHZ} we managed to determine the density of
eigenvalues for $V(\varphi)={1\over2}\varphi^2 + g
\varphi^4$ to all orders in $g$. The
correlation function was also computed, but only to first order in $g$.

This problem of ``determinism plus chance" may be
regarded
as a generic problem in physics, and as such represents a
significant generalization of Wigner's problem. For
example, consider an electron moving in a magnetic field and scattering off
impurities. We note that these ``deterministic plus random" problems
are considerably more difficult than the purely random
problems defined in \(random) and \(distribution). A standard
approach to
solving the purely random problem involves diagonalizing
the random matrix $\varphi$ and then use orthogonal
polynomials to disentangle the resulting expression. Clearly,
in \(det) we cannot diagonalize $\varphi$ without
un-diagonalizing $H_0$ and thus the orthogonal polynomial
approach
fails.

In this paper, we point out that the problem given in \(det) is
a special case of a broader class of problems involving the
addition of random matrices. The deterministic Hamiltonian
$H_0$ may in turn be replaced by a random Hamiltonian.
Indeed, a deterministic matrix is but a special case of a
random matrix. We will extend the work of Bre\'zin, Hikami,
and
Zee\refto{BHZ} and determine the density of eigenvalues
for the Hamiltonian given in \(det) for an arbitrary $V$.

Our work is inspired by recent advances in the
mathematical literature involving the theory of
non-commutative probability and operator algebra.\refto{voi,
hag, waterloo} A number of physicists have already
brought these advances to the attention of the physics
community.\refto{douglas, gross, li} While our work is thus
inspired, we will not be using the mathematical approach
given in \Ref{voi, hag, waterloo}, but instead will be based
on the diagrammatic approach developed in \Ref{BZ3} and
subsequent work.\refto{lattice, BHZ}

Consider a
Hamiltonian given by
$$
H=\varphi_1 +\varphi_2
\eqno(ham)
$$
with the matrices $\varphi_{1,2}$ taken from the
probability distribution
$$
P(\varphi_1, \varphi_2)={1 \over Z}{e^{{-N}
tr [V_1(\varphi_1) + V_2(\varphi_2)]}} \equiv P_1(\varphi_1)P_2(\varphi_2).
\eqno(distributionproduct)
$$
Notice that the probability distribution factorizes. (This is
known as ``free" in the mathematical literature.) The problem defined in
\(det) represents a special case. Previously, with D'Anna and with Br\'ezin
we have studied the problem given in \(ham) but for the
more difficult case\refto{bz4, danna} in which $
P(\varphi_1, \varphi_2)$ contains terms linking
$\varphi_1$ and $\varphi_2$. Indeed, a detailed
determination of the correlation function over all ``distance
scales" is a non-trivial problem even for Gaussian
distributions.\refto{danna} The discussion in this paper goes through
precisely because $\varphi_1$ and $\varphi_2$ do not couple to each other
in $
P(\varphi_1, \varphi_2)$.

Let us now mention a few necessary definitions.
Define the Green's function
$$
G(z)\equiv\vev{{1\over N}tr{1\over z-H}} = \int \int
d\varphi_1 d\varphi_1P(\varphi_1, \varphi_2) {{1\over N}tr{1\over
z-(\varphi_1 + \varphi_2)}}
\eqno(1.4)
$$ The density of eigenvalues is then given by
$
\rho(\mu)=\vev{{1\over N}tr\de(\mu-H)}=-{
1\over\pi} {\rm{Im}}
G(\mu+i\eps)
.$ In this paper we focus
 on the density of eigenvalues,
leaving the correlation for a future work. Note that the
factors of $N$ are chosen in our definitions such that the
interval over which $\rho(\mu)$ is non-zero is
finite (\ie of order
$N^0$) in the large $N$ limit.

We may regard the distribution \(distributionproduct) as
defining a $(0+0)$-dimensional field theory. The Feynman
diagram
expansion is then simply obtained by expanding $G(z)$ in
inverse powers of $z$ and doing the integrals in \(1.4). As explained in
\Ref{BZ3}, it is useful
to borrow the terminology of large $N$
quantum chromodynamics\refto{thoo}  from the
particle physics literature,  and speak of quark and gluon
lines. See figure (1) for a graphical representation. (It is of course not
necessary to use this language, and readers not familiar with this language
can simply think of the diagrams as representing the different terms one
encounters in doing the integral in \(1.4).) The quark propagator simply
comes from the explicit factor of $z$ in \(1.4) and is represented by a
single line and given by $1\over z$. The quadratic terms in
$V_1(\varphi_1)$ and $V_2(\varphi_2)$ determine the gluon propagator,
represented by double lines.
Here, in a minor departure from large $N$
quantum chromodynamics we
have two types of gluons, corresponding to $\varphi_1$ and
$\varphi_2$. The
gluon propagators are proportional to
$$
\vev{\varphi^i_{\alpha j}\varphi^k_{\beta l}}
 \propto \de_{\alpha \beta}\de^i_l\de^k_j{1\over N}\eqno(2.3)
$$
The non-Gaussian terms in $V_1(\varphi_1)$ and $V_2(\varphi_2)$ describe
the interaction between the gluons.

The reason that we can solve this problem is because, while the two types
of gluons
have arbitrarily complicated interactions among
themselves, they do not interact with each other. Note that
while the gluons both interact with the quark, our
problem is such that we do not have to include quark loops
and thus the quark does not induce interaction between the
two gluons. This is clear from the definition of our problem.
Another way of saying this is to note that the Green's function
may be represented, by using the replica trick, as
$$
G(z) = lim_{n \rightarrow 0} \int D\psi^{\dagger} D\psi D\varphi
P(\varphi)
\psi_1^{\dagger} \psi_1
e^{-
\sum_{\alpha=1}^n \psi_{\alpha}^{\dagger} (z-
\varphi)\psi_{\alpha}}
\eqno(field)
$$
Note that in this language the $\psi$'s represent the quark fields and
$\varphi$ the gluon fields. The interaction between gluon and quarks are
given by $ \psi_{\alpha}^{\dagger}\varphi\psi_{\alpha}$. (Color indices are
suppressed here.) The interaction of the gluons with each other is
determined by $P(\varphi)$.
Since internal quark loops are proportional to the number
of replicas $n$, they vanish in the $n\rightarrow 0$ limit.

 Let us then calculate the Green's function, which as usual can be written
as (see figure 2)
$$
G(z)={1\over z-\Si(z)}
\eqno(ipi)
$$
 in terms of the one-particle  irreducible self energy
$\Si^i_j(z)= \de^i_j \Si(z)$.  The self-energy is then
determined by the set of diagrams in figure (3) with the corresponding equation
$$\eqalign{
\Si(z)&  = <{1\over N} tr \varphi_1>_{gc}+<{1\over N} tr
\varphi_1^2>_{gc} G(z) + <{1\over N} tr \varphi_1^3>_{gc}G(z)^2 +
.... + (1 \leftrightarrow 2)\cr
& = \sum_{k=1}^{\infty} <{1\over N} tr
\varphi_1^k>_{gc}G(z)^{k-
1}  + (1 \leftrightarrow 2)\cr
&= {1\over G} [<{1\over N} tr {1\over 1-\varphi_1 G}>_{gc} -1]+ (1
\leftrightarrow 2)\cr
& = {1\over G} [{1\over G} G_{gc1}({1\over G})-1]+
(1 \leftrightarrow 2)\cr}
\eqno(cent)
$$
In order to write this equation, we have to invoke the factorization of
$P(\varphi_1, \varphi_2)$, which tells us that the two kinds of gluons do
not interact, and the large $N$ limit, which tells us that the two kinds of
gluon lines cannot cross.

We are led to
introduce in \(cent) the notion of ``gluon connectedness," denoted by ``gc"
henceforth. The necessity for this notion is illustrated by the shaded
blob describing the interaction of the gluons in figure (3d): it should not
include the diagram shown in figure (4): this class of diagrams
is already included in figure (3b). In other words, a gluon connected blob
with $k$ external gluon lines is such that it cannot be separated into two
blobs, with  $k_1$ gluon lines and $k_2$ gluon lines respectively, (with
$k_1+k_2=k$ of course). In the last line we have defined the ``gluon
connected Green's function"
$$
G_{gc1}(z)=<{1\over N} tr {1\over {z-\varphi_{1}}}>_{gc}
\eqno(ggc)
$$
and similarly
$G_{gc2}(z)$. The operations implied in \(ggc) are clearly
allowed since $< {1\over N} tr \cdot >_{gc}$ is a linear
operation. Note also that we have not assumed that $V_\alpha$ is an even
function of its argument. In particular, we include a possible
tadpole term indicated by $ <{1\over N} tr
\varphi_{\alpha}>_{gc}$ in \(cent).

We should emphasize that the shaded blobs include interactions between
gluons to all orders. It is very complicated, if not hopeless, to calculate
these blobs in terms of $V_1$ and $V_2$, but fortunately, as we will show
below, we do not have to calculate them explicitly.
In our previous papers,
we regarded the cubic, quartic, and so on, terms in
$V_\alpha$ as  interactions and proceeded to calculate the Green's
function and correlation function in terms of
the various coupling constants. We follow a different
strategy here, and try to express the Green's function
$G(z)$ directly in terms of $G_1(z)$ and $G_2(z)$ where
$$
G_{\alpha}(z) \equiv <{1\over N} tr {1\over  z -
\varphi_{\alpha} }>
\eqno(alpha)
$$
for $\alpha =1,2$ are the Green's functions for two
separately and purely random problems. (The average in \(alpha)
is performed with the distribution $P_\alpha(\varphi_\alpha)={1\over
Z_\alpha}e^{-Ntr
V_\alpha(\varphi_\alpha)}$ of course.) In this way, we
attempt to bypass having to deal with $V_1$ and $V_2$
altogether.

To see how to do this, let us go back to the simpler problem defined by
\(random) and
\(distribution).
Following the same diagrammatic analysis leading to\(cent) we find that the
Green's function $G(z)$ and self energy $\Si(z)$ for this simpler problem
are related by
$$
\Si(z) = {1\over G} [{1\over G} G_{gc}({1\over G})-1]
\eqno(cent1)
$$
with, evidently,
$$
G_{gc}(z) \equiv <{1\over N} tr {1\over {z-\varphi}}>_{gc}
\eqno(def)
$$
Combining \(def) and \(ipi) we find
$$
{1\over G^2} G_{gc}({1\over G}) = z
\eqno(zip)
$$
It is convenient to define a ``Blue's function"
$$
B(z) \equiv {1 \over z^2} G_{gc}({1 \over z})
\eqno(blue)
$$
Thus, we learn that the Blue's function is the functional
inverse of the
Green's function
$$
B(G(z))=z
\eqno(inverse)
$$
 From the normalization of the probability distribution
$P(\varphi)$ we
obtain trivially the ``sum rule" $G(z)\rightarrow{1\over z}$
as $z\rightarrow\infty$, thus
implying that the Blue's function $B(z)\rightarrow{1\over
z}$ as $z\rightarrow0$.

Let us now go back to the more involved problem
defined by \(ham). First, we define for $\alpha=1,2$ two Blue's functions
$B_{\alpha}$ as the
functional inverse
of $G_{\alpha}$ respectively. We now see that \(cent), when combined with
\(ipi), says simply that
$$
z+{1\over G}= B_1(G) + B_2(G)
\eqno(add)
$$
Thus, the law of addition for the Blue's function is given by
$$
B_{1+2}(z)=B_1(z)+B_2(z)-{1\over z}
\eqno(addlaw)
$$
This equation tells us how to obtain the Blue's function associated with
$\varphi_1 +
\varphi_2$ from the Blue's functions associated with
$\varphi_1$ and
$\varphi_2$.

The procedure for determining the Green's function and
hence the density of
eigenvalues of the problem defined by \(random) and
\(distribution) is then
as follows: given the Green's functions $G_1$ and $G_2$,
determine the
corresponding Blue's functions $B_1$ and $B_2$,  calculate
$B_{1+2}$
according to \(addlaw), then determine the functional
inverse of  $B_{1+2}$
to find the desired Green's function $G(z)$.

Let us remark briefly on the connection to the mathematical
literature.
Voiculescu\refto{voi} has introduced the ``$R$-transform." It
turns out that
the $R$ function discussed by mathematicians is simply
related to $B$ by
$B(z)={1\over z} + R(z)$. In fact, we see that the
Dyson-Schwinger equation
\(ipi) when combined with \(inverse) gives simple
$B(G(z))={1\over G(z)} +
\Si(z)$. Thus, the $R$ function of the mathematicians is
nothing but the
self-energy $\Si$ of the physicists expressed in terms of
different
arguments:
$$
R(G(z))=\Si(z)
\eqno(math=phys)
$$

Let us now proceed by building up from a few simple
examples. In the most trivial case, $\varphi$ is
not random at all, but fixed to be a constant $c$ times the
unit matrix. Then from the Green's function $G(z)= {1\over
z-c}$ we find the Blue's function $B(z)=c+{1\over z}$. For a
slightly less trivial example, let $\varphi$ be a diagonal
matrix with matrix elements given by $\epsilon_i$ with $i=1,
..., N$. Then the corresponding Blue's function is determined
by
$$
{1\over N}\sum_i{1\over B(z)-\epsilon_i}=z
\eqno(bluedet)
$$
Next, let $P(\varphi)$ be  Gaussian (that is, $V(\varphi)=tr
{1\over2}\varphi^2$). Then as is well known (see for example \Ref{BZ3}), the
Green's function is determined by
$$
z=G(z)+{1\over G(z)}
\eqno(gauss)
$$
Substituting $z\rightarrow B$, we obtain immediately that\refto{addgauss}
$$
B(z)=z+{1\over z}
\eqno(gaussb)
$$

Now we are ready to do our first non-trivial problem.
Consider the problem defined in \(det). Since we know from
\(bluedet) and \(gaussb) the Blue's functions corresponding
to the two terms in the Hamiltonian, we learn immediately
from \(addlaw) the Blue's function for $H$:
$$
B_{1+2}(z)=B_1(z)+z+{1\over z}-{1\over z}
=B_1(z) + z
\eqno(fullblue)
$$
with $B_1$ determined by \(bluedet) with the substitution
$B \rightarrow B_1$.
The desired Green's function $G(z)$ is now determined by
solving for the functional inverse
$$
B_{1+2}(G)=z
\eqno(iii)
$$
or equivalently, upon using \(fullblue)
$$
B_1(G)=z - G
\eqno(iv)
$$
Let us now evaluate the two sides of this equation with the
function $G_1(\cdot)$. We obtain immediately
$$
G(z)=G_1(z-G(z))
\eqno(past)
$$
precisely the classic result of Pastur which was obtained
diagrammatically in \Ref{BZ3}.

After these simple exercises, we can now immediately go on
and solve the general version of the problem defined in
\(det): find the density of energy levels of a Hamiltonian
given by $H=H_0 + \varphi$ with $\varphi$ drawn from the
general distribution \(distribution). With a slight shift in
notation, let us call the Blue's function associated with $H_0$
and with $\varphi$ respectively $B_0$ and $B_2$. Then the
Blue's function associated with $H$ is given by $B(z)=B_0(z)
+B_2(z) - {1\over z}$. Substituting $z \rightarrow G(z)$
(where $G(z)$ is the unknown Green's function associated
with $H$), we find immediately that
$$
B_0(G)=z+{1\over G}-B_2(G)
\eqno(daf)
$$
Anticipating the next step, we define
$$\Si(z)=B_2(G(z))-{1\over G(z)}
\eqno(adef)
$$
Let us now evaluate both sides of \(daf) with the Green's
function $G_0(\cdot)$ associated with $H_0$. We find instantly
that
$$
G(z)=G_0(z-\Si(z))
= {1\over N} \sum_i {1\over z-\epsilon_i-\Si(z)}
\eqno(one)
$$
Let us repeat this trick: rewrite \(adef) as
$B_2(G(z))=\Si(z)+{1\over G(z)}$ and evaluate both sides
with the function $G_2(\cdot)$. We obtain
$$
G(z)=G_2(\Si(z) + {1\over G(z)})
\eqno(other)
$$

These two equations, \(one) and \(other), allow us to
determine the two unknown functions $G(z)$ and $\Si(z)$,
provided we know the Green's function $G_2(z)$. But what
is the Green's function $G_2(z)$? It is just the Green's
function associated with the random matrix $\varphi$
drawn from the general distribution \(distribution). But this was obtained
by Br\'ezin et al\refto{BIPZ} almost
twenty years ago. These authors told us that (for $V(z)$ an
even polynomial for the sake of notational simplicity)
$$
G_2(z)={1\over2}[V'(z)-P(z){\sqrt {z^2 - a^2}}]
\eqno(bipz)
$$
Here $V'(z)\equiv {dV \over dz}$, $P(z)$ is a polynomial, and
$a$ determines the endpoints of the spectrum of
eigenvalues. The quantities $P(z)$ and $a$ are determined\refto{foot}
by the ``sum rule" that $G_2 \rightarrow {1\over z}$ as $z
\rightarrow \infty$.

Thus, in summary, for any distribution defined by $V(\varphi)$, we can
determine the Green's function $G(z)$ and hence the density of
eigenvalues by solving \(one) and \(other). It is clearly convenient
to define $\si(z) \equiv \Si(z) + {1\over G(z)}$. We can then simplify
\(other) slightly to\refto{passing}
$$
P^2(\si)(\si^2 - a^2)=(V'(\si)-2G)^2
\eqno(sigma)
$$
Thus, we can use \(sigma) to determine $\si$, and hence $\Si$, in terms
of $G$. Plugging this into \(one) then gives us an equation for $G$.

As mentioned earlier, Br\'ezin, Hikami,
and Zee\refto{BHZ} recently used the equation of motion method and a detailed
diagrammatic analysis to determine  the Green's function for the
problem in
\(det) with the distribution defined by $V(\varphi)={1\over
2}\varphi^2 + g \varphi^4$. It is straightforward, although slightly
tedious, to verify that for this simple
case, \(sigma) reduces to equation (4.15) in \Ref{BHZ}. The analysis
given here is considerably simpler.

In practice, with an arbitrary set of $\epsilon_i$, \(one) can only be
solved numerically. One relatively simple example involves
taking half of the $\epsilon_i$'s to be equal to $+\epsilon$ and the
other half to be equal to $-\epsilon$. Thus, before the introduction
of the random term $\varphi$ into the Hamiltonian, the spectrum
consists of two levels, which may for example represent the two
lowest Landau levels, corresponding to spin up and spin down, in
a spin-dependent quantum Hall fluid.\refto{hikami, girvin, hz}.
The randomness will then broaden the two levels.

Clearly, by repeating the discussion given here, we can add an
arbitrary number of random Hamiltonians together. The
procedure is defined by \(addlaw). The deterministic plus random
Hamiltonian studied here is just a special case. It will be
interesting to see if this work can be generalized to a study of the
universal correlation function discussed in \Ref{BZ1, BZ3, BHZ}.

\head{Acknowledgement}

I thank the organizers of the workshop on ``Operator Algebra Free Products
and Random Matrices" held at the Fields Insitute, March 1995, for inviting
me to talk about my work. I had the opportunity there to learn  from
mathematicians, in particular U. Haagerup and D. Voiculescu, about recent
developments. I am especially grateful to M.  Douglas for many instructive
discussions and for telling me how to translate from the language used by
mathematicians to the language used by physicists. Finally, I would like to
thank E. Br\'ezin, my collaborator over the last few years, for countless
helpful discussions about this and other topics in random matrix theory and
for reading this manucsript.
This work was supported in part by the National Science
Foundation under Grant No. PHY89-04035.

\head{Figure Captions}

Fig 1. Feynman rules: (a) quark propagator, (b) gluon propagator, (c) quark
gluon vertex, (d) gluon interaction, illustrated here with a $g \varphi^4$
vertex.

Fig 2. Quark propagator and one-particle-irreducible self energy.

Fig 3. Quark self energy: the gluons shown explicitly are all of type 1.
There are of course also type 2 gluons inside the quark propagator $G$.

Fig 4. A class of diagrams not included in (3d).

\references

\refis{BZ1} E. Br\'ezin and A. Zee, \np 402(FS), 613, 1993.

\refis{BZ2} E. Br\'ezin and A. Zee, {\sl Compt.\ Rend.\ Acad.\
Sci.\/}
(Paris) t.317 II, 735 (1993).

\refis{BZ3} E. Br\'ezin and A. Zee, Phys. Rev. {\bf E49} (1994) 2588.

\refis{lattice} E. Br\'ezin and A. Zee, \np B441(FS), 409, 1995.

\refis{bz4} E. Br\'ezin and A. Zee, \np B424(FS), 435, 1994.

\refis{WIG} E. Wigner, {\sl Can.\ Math.\ Congr.\ Proc.\/}
p.174
(University of
Toronto Press) and other papers reprinted in Porter, op. cit.

\refis{POR} C.E. Porter, {\it Statistical\ Theories\ of
 \
Spectra:\ \
Fluctuations\/}
(Academic Press, New York, 1965).

\refis{girvin} C. B. Hanna, D. P. Arovas, K. Mullen, and S. M. Girvin,
Indiana preprint (1994). cond-mat.9412102.

\refis{MEH} M.L. Mehta, {\it Random\ Matrices\/} (Academic
Press, New
York,
1991).

\refis{JER} See for instance, {\it Two\ Dimensional\ Quantum\ Gravity\ and\
Random\ Surfaces\/}, edited by D.J.
Gross and T.
Piran (World Scientific, Singapore, 1992).


\refis{BHZ} E. Br\'ezin, S. Hikami and A. Zee,
           Paris-Tokyo-Santa Barbara preprint (1994) LPENS-94-35,
NSF-ITP-94-135, UT-
KOMABA-94-21. (hep-th.9412230).

\refis{hikami} S. Hikami, M. Shirai, and F. Wegner, Nucl. Phys. {\bf
B408} (1993) 415.

\refis{hz} S. Hikami and A. Zee, Tokyo-Santa Barbara preprint
(1995), cond-mat/9504014, {\sl Nucl. Phys.} FS, in print,
1995.

\refis{danna} J. D'Anna, E. Br\'ezin, and A. Zee, {\sl Nucl. Phys.} FS, in
print,
1994.

\refis{gross} R. Gopakumar and D. J. Gross,  Princeton preprint PUPT-1520,
1994.

\refis{douglas} M. Douglas, Rutgers preprint, hep-th/9409098, 1994.

\refis{li} M. Douglas and M. Li, Rutgers preprint, 1995.

\refis{voi} D. V. Voiculescu, K. J. Dykema, and A. Nica, {\it Free\ Random\
Variables\/} (AMS, Providence, R. I.,
1992).

\refis{hag} U. Haagerup, private communication and to be published.

\refis{waterloo} Lectures at the workshop on ``Operator Algebra Free
Products and Random Matrices," Fields Insitute, March 1995, to appear in
the Proceedings.

\refis{bee} C.W.J. Beenakker,  \np B422(FS), 515, 1994.

\refis{eyn}   B. Eynard,  {\sl Nucl. Phys.} FS, in print,
1994.

\refis{for} P. J. Forrester, \np B435(FS), 421, 1995.

\refis{thoo} G. 't Hooft, \np B72, 461, 1974.

\refis{BIPZ} E. Br\'ezin, C. Itzykson, G. Parisi, and J.B. Zuber,
\journal  Comm. Math. Phys., 59, 35, 1978.

\refis{PAS} L.A. Pastur, \journal Theo. Math. Phys., 10, 67,
1972.

\refis{koba} T. S. Kobayakawa, Y. Hatsugai, M. Kohmoto, and A. Zee,
ISSP-Tokyo preprint (1994)

\refis{foot} For the sake of completeness, let us record (and for the sake of
simplicity, with $V$ an even function) that
for $V(\varphi)=\sum_{k=1}^p {1\over 2k}g_k \varphi^{2k}$ we have
$$
P(z)={1\over 2}\sum_{k=1}^p g_k \sum_{n=0}^{k-1} {(2n)!\over (n!)^2}
({a^2\over 4})^n \lambda^{2k-2n-2}
$$    and
$$
{1\over 2}\sum_{k=1}^p g_k {(2k)!\over (k!)^2}({a^2\over 4})^k =1
$$
In particular, for $V={1\over 2}\varphi^2+{g\over 4}\varphi^4$ we have
$$
G(z)={1\over 2}[z+gz^3-(1+{a^2 g\over 2}+g z^2) {\sqrt {z^2-a^2}}]$$ and
$a^2={2\over 3g}(\sqrt {1+12g}-1)=4(1-3g+18g^2+....)$.

\refis{passing} It is perhaps worth noting that $B_2$ satisfies an equation
similar to \(sigma), namely
$$
P^2(B_2)(B_2^2 - a^2)=(V'(B_2)-2z)^2
$$

\refis{addgauss} Using dimensional analysis we see immediately that for
$V(\varphi)={1\over 2}m^2 \varphi^2$ the Blue's function $B(z)={z \over
m^2}+{1\over z}$. Then \(fullblue) implies the usual Gaussian law of
addition $m^{-2}_{1+2}=m^{-2}_1 + m^{-2}_2$.

\endreferences

\end